\magnification=1200
\null\vskip1.5cm
\centerline{\bf SELF-DUAL VORTICES IN CHERN-SIMONS HYDRODYNAMICS}
\bigskip\medskip
\centerline {Oktay K. PASHAEV$^{1,2}$ and Jyh-Hao LEE$^3$ }
\bigskip
\centerline
{\sl $^1$Department of Mathematics, Izmir Institute of Technology,
Cankaya-Izmir, Turkey}
\smallskip
\centerline
{\sl $^2$Joint Institute for Nuclear Research, Dubna (Moscow),
Russian Federation}
\par\smallskip
\centerline
{\sl $^3$Institute of Mathematics, Academia Sinica, Taipei, Taiwan}

\par\smallskip
\bigskip
\centerline{Abstract}
\medskip
{\sl The classical theory of non-relativistic charged particle
interacting with $U(1)$ gauge field is reformulated as the
Schr\"odinger wave equation modified by the de-Broglie-Bohm quantum potential
nonlinearity. For, (1 - $\hbar^2$)
deformed strength of quantum potential
the model is gauge equivalent to the standard Schr\"odinger equation
with Planck constant $\hbar$,
while for the strength (1 + $\hbar^2$), to the pair of
diffusion-anti-diffusion equations.
Specifying the gauge field as Abelian Chern-Simons (CS) one in 2+1 dimensions
interacting with
the Nonlinear Schr\"odinger field
(the Jackiw-Pi model), we represent the theory
as a planar Madelung fluid, where the Chern-Simons Gauss
law has simple physical meaning of creation
the local
vorticity for the fluid flow. For the static flow,
when velocity of the center-of-mass motion (the classical velocity)
is equal to the quantum one (generated by quantum potential velocity of
the internal motion),
the fluid admits N-vortex solution.
Applying the Auberson-Sabatier type gauge
transform to phase of the vortex wave function we show that
deformation parameter $\hbar$,
the CS
coupling constant and the quantum potential strength are quantized.
Reductions of the model to 1+1 dimensions, leading to modified
NLS and DNLS
equations with resonance soliton interactions are discussed.}
\bigskip
\noindent {\bf 1. Introduction}
\bigskip\par
Nonlinear extension of the Schr\"odinger equation by the "quantum potential"
non-linear term has been considered long time ago in connection with
a stochastic quantization problem
[1], and corrections to quantum mechanics from quantum gravity [2].
It appears also in the wave theoretical formulation of classical mechanics [3],
and the dispersionless limit of nonlinear wave dynamics [4].
As was shown by Sabatier [5]
this extension preserves the Lagrangian structure. Moreover,
by proper transformation of the wave function's phase, Auberson and
Sabatier [6] obtained linearization of the model, which depending on the
strength of the quantum potential, has to appear in the form of the
Schr\"odinger equation with rescaled potential or as the pair of the
time-reversed diffusion equations. Due to this linearization no
soliton type solutions were found [5,6]. Meanwhile,
recently we have been considering the nonlinear version of the
Bohm's formulation of the quantum mechanics [7], namely
the problem of the Nonlinear Schrodinger (NLS) soliton under the
influence of the quantum potential [4,8].
Application of
the Auberson-Sabatier type phase transform to
this problem
with over-critical strength of the quantum potential,
allowed us reduce the problem to the pair of time-reversed reaction-diffusion
equations, representing an imaginary time version of the real q-r NLS
type system [9] (SL(2,R) reduction of the Zhakharov-Shabat problem).
Then, constructing two soliton solution we found a resonance character
of their mutual interaction [4,8].
\par
In the present paper we consider the influence of quantum potential
on the planar vortex in 2+1 dimensional problem for the Nonlinear
Schr\"odinger equation interacting with the Chern-Simons gauge field.
Application of the Auberson-Sabatier type transform, affecting the phase
of the wave function, dramatically changes parameters of the vortex
configurations. In the Madelung representation we reformulate
the model as a rotational planar hydrodynamics. Then,
the self-dual limit, admitting N-vortex solutions, has simple physical
interpretation as a condition of equality between "classical velocity"
(velocity of the center-of-mass) and the "quantum" one (velocity of the
motion in the center-of-mass frame associated  with the internal
"spin motion" or {\it zitterbewegung}).
\par
In Section 2 we reformulate the classical dynamics of charged particle
interacting with Abelian gauge field as a nonlinear Schr\"odinger type
wave equation. Deforming properly the strength of quantum potential
we recover the standard Schr\"odinger equation, where the deformation
parameter plays role of the Planck constant. In Section 3 we specify
the gauge field as an Abelian Chern-Simons one,
 interacting with the Nonlinear Schr\"odinger
equation (NLS), and derive corresponding rotational Madelung type hydrodynamics,
its dispersionless limit and deformations.
Section 4 devoted to the quantum velocity and its properties. For the static
flow moving with a velocity equal to the quantum one
 we reduce the problem to the Liouville
equation and describe corresponding vortex configurations. From conditions
of non-singularity and single valuedness we find quantization condition
for coupling constants. Dimensional reduction to one dimensional
NLS equation and its modification by quantum potential are considered
in Section 5. In Conclusions we briefly discuss our results.
\par\bigskip
\noindent{\bf 2. Nonlinear Wave Equation of Classical Dynamics  }
\bigskip\par
The classical dynamics of charged non-relativistic particle
in U(1) gauge field
$A_{\mu} = (A_{0}, {\bf A} )$,
with
the Hamiltonian function
$$H = {{\bf p}^2 \over 2 m} + {e \over c}A_{0} + U, \eqno(2.1)$$
is described by the Hamilton-Jacobi equation
$${\partial S\over \partial t} + H (\nabla S, A_{0}, {\bf A}, U) = 0,
\eqno(2.2)$$
where for the momentum ${\bf p}$ we substitute
$${\bf p} = \nabla S + {e \over c} {\bf A}.\eqno(2.3)$$
Combining (2.2) with the Liouville equation
$${\partial \rho \over \partial t} + \nabla (\rho {\bf V}) = 0,
\eqno(2.4)$$
for the density $\rho$
of integral invariant in the gradient dynamical system
$$\dot{\bf x} =  {\bf V} = {1 \over m}{\bf p} =
{1 \over m} [\nabla S + {e \over c} {\bf A}],
\eqno(2.5)$$
we have the system of equations
$$\cases {{\partial S\over \partial t} +
{1 \over 2m}(\nabla S + {e \over c}{\bf A})^2
+ {e \over c} A_{0} + U = 0, \cr\cr
{\partial \rho \over \partial t} + \nabla (\rho {\bf V}) =
0.  \cr}    \eqno(2.6)$$
This classical system is representable in the wave form.
Introducing the complex wave function ("order parameter")
$$\psi = \sqrt{\rho}e^{i S}\,\,,\eqno(2.7)$$
we rewrite  equations (2.6) as the
single nonlinear wave equation
$$ i D_{0}\psi + {1 \over 2m}{\bf D}^2 \psi - U \psi
= {1 \over 2 m}{\Delta |\psi| \over |\psi|} \psi \,,\eqno(2.8)$$
where $D_{0} = \partial_{t} + {e \over c}A_{0}$,
      ${\bf D} = \nabla + {e \over c} {\bf A}$.
The last equation has form of the Schr\"odinger equation (without any
Planck constant) modified by the
so called "quantum potential" term in the right hand side.
It admits all the usual solutions
of classical mechanics
but does not allow superpositions of these solutions.
Since the system (2.6) describes the formal semiclassical limit
of the quantum mechanical Schr\"odinger equation, Eq. (2.8) can be considered
as its dispersionless
limit.
In fact,
the wave equation (2.8) is covariant under the gauge transformations
$$\psi \rightarrow \psi e^{i \alpha},\,\,\,
{\bf A} \rightarrow {\bf A} - {c \over e}\nabla \alpha,\eqno(2.9)$$
generating shift of the classical action
$$S \rightarrow S + \alpha.$$
Then, being U(1) gauge invariant,
the additional term on the r.h.s. of (2.8) completely compensates
the corresponding gauge invariant contribution from
dispersion in the l.h.s.
\par
As we mentioned above, Eq.(2.8) does not contain the Planck constant.
But if we consider contribution from
quantum potential in the r.h.s.  of Eq. (2.8) deformed
by a constant $\hbar^2$
  $$ i D_{0}\psi + {1 \over 2m}{\bf D}^2 \psi - U \psi
= {1 \over 2 m}(1 - \hbar^2){\Delta |\psi| \over |\psi|} \psi\,, \eqno(2.10)$$
then in terms of the new wave function
$$\chi = \sqrt{\rho} e^{{i\over \hbar} S},\eqno(2.11)$$
we recover the standard linear Schr\"odinger equation
 $$ i \hbar D_{0}\chi + {\hbar^2 \over 2m}{\bf D}^2 \chi - U \chi
= 0,\eqno(2.12)$$
where $\hbar$ plays the role of the Planck constant.
Thus
for $\hbar \neq 0$
Eq.(2.10) is gauge equivalent to the Schr\"odinger equation and
for $\hbar = 0 $ reduces to the nonlinear wave equation
of classical mechanics
(2.8). Moreover, for $\hbar = \pm 1$ it reduces directly
to the linear Schr\"odinger
equation and its complex conjugation. From another site
if the deformation of Eq.(2.8) appears with an opposite sign as
  $$ i D_{0}\psi + {1 \over 2m}{\bf D}^2 \psi - U \psi
= {1 \over 2 m}(1 +  \hbar^2){\Delta |\psi| \over |\psi|} \psi \,,\eqno(2.13)$$
nevertheless to the same classical limit $\hbar = 0$ as for Eq.(2.10), it
cannot be  linearized in the form
of the Schr\"odinger equation by transformation (2.11). However,
if we notice that Eq.(2.13) can be reduced to Eq.(2.10) by
formal analytical substitution for the Planck constant
to the pure imaginary value $\hbar \rightarrow i \hbar$
(in quantum mechanics similar continuation to
the classically inaccessible region leads to  the
exponentially decaying (growing) wave function). Then,
written in terms of two real
functions
$$Q^{\pm} = \sqrt{\rho} e^{\pm {1 \over \hbar}S}\,,\eqno(2.14)$$
Eq.(2.13) and its complex conjugation
become the pair of decoupled diffusion-antidiffusion
equations
 $$ \pm \hbar  D_{0}Q^{\pm} + {\hbar^2 \over 2m}{\bf D}^2 Q^{\pm}
- U Q^{\pm} = 0,\eqno(2.15)$$
similar to the one considered  by Schr\"odinger in 1931 [10].
From the above consideration we see that Schr\"odinger
equation perturbed by quantum
potential includes as a particular cases
the classical mechanics ($\hbar = 0$), the
quantum mechanics ($\hbar = \pm |\hbar|$) and
the pair of diffusion-antidiffusion equations
($\hbar = i|\hbar|$).
\par

\bigskip
\noindent{\bf 3. Chern-Simons Hydrodynamics }
\bigskip\par
The semiclassical limit has been applied recently to
defocusing Nonlinear Schr\"odinger (NLS) equation
$$i\hbar \partial_{t}\chi + {\hbar^2 \over 2 m}\Delta \chi
+ 2 g |\chi|^2 \chi = 0\,,\eqno(3.1)$$
($g < 0$) in one [11] and two space dimensions [12] and provides an analytical
tool to describe shock waves in nonlinear optics and
vortices in superfluid. Decomposing the wave function like in
(2.11) one derives quantum deformation of the Hamilton-Jacobi
equation by quantum potential, or after differentiation according to
space coordinates, the Madelung fluid. In the formal
semiclassical limit $\hbar \rightarrow 0$ (before shocks appear),
one neglects
contribution from the quantum potential and fluid becomes the
Euler system. Then in terms of the wave function (2.7)
we have dispersionless NLS equation
$$i\partial_{t}\psi + {1 \over 2 m}\Delta \psi
+ 2 g |\psi|^2 \psi = {1 \over 2 m}{\Delta |\psi|\over |\psi|}\psi\,.
\eqno(3.2)$$
The  quantum deformation of the last equation in the form
$$i\partial_{t}\psi + {1 \over 2 m}\Delta \psi
+ 2 g |\psi|^2 \psi = (1 - \hbar^2)
{1 \over 2 m}{\Delta |\psi|\over |\psi|}\psi\,,\eqno(3.3)$$
reformulated for the wave function (2.11), leads us again to the original equation (3.1).
\par
The NLS model (3.1) interacting with  Chern-Simons
gauge field in 2+1 dimensions is called the Jackiw-Pi (JP)
model and describes anyonic phenomena [13]. The semiclassical limit of
anyons requires to study this model in the limit when $\hbar \rightarrow 0$,
or similarly to the case of Eq. (3.2),  its perturbations by quantum potential.
\par
To describe the deformed theory
we consider the Lagrangian
$$L = {\kappa \over 2}\epsilon^{\mu\nu\lambda}
A_{\mu}\partial_{\nu}A_{\lambda} + {i \over 2}(\bar\psi D_{0} \psi
- \psi \bar D_{0}\bar \psi) - {1 \over 2 m} |{\bf D}\psi|^2
+ (1 - \hbar^2) {1 \over 2 m}(\nabla |\psi|)^2
+ g |\psi|^4, \eqno(3.4)$$
where $D_{\mu} = \partial_{\mu} + {i e \over c}A_{\mu}$,
leading to the system of equations of motion
$$ i D_{0}\psi + {1 \over 2m}{\bf D}^2 \psi + 2g |\psi|^2 \psi
= (1 - \hbar^2) {1 \over 2 m}{\Delta |\psi| \over |\psi|} \psi\,, \eqno(3.5a)$$
$$\partial_{1} A_{2} - \partial_{2} A_{1} =
{e \over \kappa c}\bar\psi \psi \,,\eqno(3.5b)$$
$$\partial_{0} A_{j} - \partial_{j}  A_{0} = - {i e \over 2 mc \kappa}
\epsilon_{jk} (\bar\psi D_{k}\psi - \psi \bar D_{k}\bar \psi) \,.\eqno(3.5c)$$
Decomposing the wave function $\psi = \sqrt{\rho} \exp (iS)$ as in Eq.(2.7),
and introducing the new function $\chi = \sqrt{\rho} \exp ({iS/\hbar})$
as in Eq.(2.11),
we have the Jackiw-Pi model
$$ i \hbar (\partial_{0} + {ie \over \hbar c}A_{0})\chi +
{\hbar^2 \over 2m}(\nabla +
{ie \over \hbar c} {\bf A})^2 \chi + 2g |\chi|^2 \chi
= 0\,, \eqno(3.6a)$$
$$\partial_{1} A_{2} - \partial_{2} A_{1} =
{e \over \kappa c}\bar\chi \chi\,, \eqno(3.6b)$$
$$\partial_{0} A_{j} - \partial_{j}  A_{0} = - {i e \hbar \over 2 mc \kappa}
\epsilon_{jk} [\bar\chi (\partial_{k} +
{ie\over \hbar c}A_{k})\chi - \chi (\partial_{k}
- {ie \over \hbar c} A_{k})\bar \chi]\,. \eqno(3.6c)$$
Corresponding Lagrangian follows from (3.4) as
$$L = {\kappa \over 2}\epsilon^{\mu\nu\lambda}
A_{\mu}\partial_{\nu}A_{\lambda} + $$
$${i\hbar \over 2}[\bar\chi (\partial_{0} +
{ie \over \hbar c}A_{0}) \chi
- \chi (\partial_{0} - {ie\over \hbar c}A_{0})\bar \chi] -
{\hbar^2 \over 2 m}
(\nabla - {ie\over \hbar c}{\bf A})\bar\chi (\nabla +
{ie\over \hbar c}{\bf A})\chi
+ g |\chi|^4\,. \eqno(3.7)$$
In the above system of equations (3.6) the deformation parameter $\hbar$ plays the role
similar to the Planck constant.
Both of the systems (3.5) and (3.6)
admit the same hydrodynamical (Madelung type)
representation.
From equation (3.5a)
we obtain the quantum Hamilton-Jacobi equation
$${\partial S \over \partial t} + [{m V^2 \over 2} +
{e \over c}A_{0} - 2g \rho
- {\hbar^2 \over 2 m} {\Delta \sqrt{\rho} \over \sqrt{\rho}}] = 0\,,
\eqno(3.8)$$
and the continuity one
$${\partial \rho \over \partial t} + \nabla (\rho {\bf V}) = 0 \,,\eqno(3.9)$$
where like in Eq.(2.5) we introduced the local velocity field
$${\bf V} = {1 \over m} [\nabla S + {e \over c} {\bf A}]\,.\eqno(3.10)$$
Then equations (3.6b,c) become of the form
$$\partial_{1} A_{2} - \partial_{2} A_{1} =
{e \over \kappa c}\rho, \eqno(3.11)$$
$$\partial_{0} A_{j} - \partial_{j}  A_{0} = {e \over \kappa c}
\epsilon_{jk} \rho V_{k}. \eqno(3.12)$$
Now we can completely exclude from consideration the
vector potentials ${\bf A}$
in favor of the velocity field (3.10). It is worth to note
that the last one is an explicitly gauge invariant variable.
Thus from Eqs.(3.12) and (3.8) we
derive the Euler equation for velocity ${\bf V}$,
$${\partial {\bf V} \over \partial t} + ({\bf V} \nabla) {\bf V}
= - {1 \over m}\nabla P\,,\eqno(3.13)$$
with the pressure
$$P =  - 2g \rho
- {\hbar^2 \over 2 m} {\Delta \sqrt{\rho} \over \sqrt{\rho}}\,.
\eqno(3.14)$$
The Chern-Simons Gauss law (3.11) in terms of our hydrodynamical variables
becomes
$$\nabla \times {\bf V} = {e^2 \over m \kappa c^2}\, \rho\,. \eqno(3.15)$$
The last condition has simple meaning of a rotational fluid, such that at any point
of the fluid
with nonvanishing density $\rho$ the local vorticity is nonzero.
The system of equations (3.9),(3.13-15) determines
the Madelung fluid for our model.
As well as the velocity field mentioned above in (3.10)
it is  explicitly $U(1)$ gauge invariant.
Moreover the continuity equation (3.9) from this system is not
independent. It appears as a consistency condition for the
Chern-Simons Gauss law (3.15) during the time evolution.
To check this it is sufficient simply differentiate (3.15)
according to the time variable and use Eq.(3.13).
Thus we have
hydrodynamical model defined by two equations
$$\cases{{\partial {\bf V} \over \partial t} + ({\bf V} \nabla) {\bf V}
= - {1 \over m}\nabla (- 2g \rho
- {\hbar^2 \over 2 m} {\Delta \sqrt{\rho} \over \sqrt{\rho}})\,,
 \cr
\nabla \times {\bf V} = {e^2 \over m \kappa c^2} \rho\,. \cr}\eqno(3.16)$$
The semiclassical or dispersionless limit of this model when
$\hbar \rightarrow 0$, is given by
$$\cases{{\partial {\bf V} \over \partial t} + ({\bf V} \nabla) {\bf V}
= - {1 \over m}\nabla (- 2g \rho)\,,\cr
\nabla \times {\bf V} = {e^2 \over m \kappa c^2} \rho \,.\cr}\eqno(3.17)$$
The nonlinear wave form of these equations follows directly from
the system (3.5) and Lagrangian from Eq.(3.4) with $\hbar = 0$.
\bigskip
\par\noindent
{\bf 4. Quantum velocity and stationary flow}
\bigskip\par
Recently in the set of papers, an interpretation
of quantum potential in terms of velocity of internal motion or the
{\it zitterbewegung} was given [14]. In that  approach
starting from the Pauli
current a decomposition of the nonrelativistic
local velocity in two parts,
one parallel and the other orthogonal to the impulse has been obtained.
The first part, determined by $\nabla S$,
is recognized as the "classical" part,
corresponding to the velocity of the center-of-mass. While the second one,
called the "quantum" one, is the velocity of motion
in the center-of-mass frame
(the internal "spin motion" or Schr\"odinger's {\it zitterbewegung}).
Then the contribution of the quantum potential
to the Lagrangian (3.7) has
simple physical meaning of the kinetic energy for this last motion
$${\hbar^2 \over 2 m}(\nabla |\chi|)^2 =
{\hbar^2 \over 8 m}({\nabla \rho \over \rho})^2
= {{m {\bf V}^2_{q}} \over 2}\,, \eqno(4.1)$$
where the "quantum" velocity is defined by
$${\bf V}_{q} = {\nabla \rho \times {\bf s} \over m \rho}\,.\eqno(4.2)$$
For the planar motion in the $x-y$ plane, $\nabla_{z} = 0$
and $s_{x} = s_{y} = 0$, so that $s_{z} = {\hbar \over 2}$, and
we have for components of quantum velocity
$$(V_{q})_{x} = {\hbar \over 2m} {\partial_{y} \rho \over \rho},\,\,\,
     (V_{q})_{y} = -{\hbar \over 2m} {\partial_{x} \rho \over \rho},
\eqno(4.3)$$
or
$$ ({\bf V}_{q})_{i} = {\hbar \over 2m} \epsilon_{ij}
{\partial_{j} \rho \over \rho}\,.
\eqno(4.4)$$
Differentiating in time the last equation and using the continuity
equation (3.9)  we obtain
$$\partial_{0} {\bf V}_{q} + ({\bf V}\nabla) {\bf V}_{q} = 0,\eqno(4.5)$$
which means that  ${\bf V}_{q}$ is propagating with the
main flow velocity ${\bf V}$, i.e. it is a velocity of the inner
motion. Moreover, by direct computation from Eq.(4.4)
we have divergenceless
condition for the quantum velocity flow
$$\nabla (\rho {\bf V}_{q}) = 0.\eqno(4.6)$$
This condition applied to the continuity equation (3.9)
for a flow propagating with quantum velocity
$${\bf V}= \pm{\bf V}_{q},\eqno(4.7)$$
having meaning of
a special planar
motion when velocities of classical (center-of-mass) motion and
quantum (internal)
motion coincide,  leads to the  stationary flow
$$\partial_{0}\rho = 0 \,.\eqno(4.8).$$
But from another site for the stationary flow when
$$\partial_{0}{\bf V} = 0\,,\eqno(4.9)$$
and
$${\kappa g \over e^2} = \pm{\hbar \over 2 mc^2}\,,\eqno(4.10)$$
the Madelung fluid equation (3.16) can be rewritten as
$${m \over 2}\nabla_{j} ({\bf V} - {\bf V}_{q})({\bf V} + {\bf V}_{q})
- {e^2 \over \kappa c^2}\rho \epsilon_{jk}({\bf V} \mp {\bf V}_{q})_{k}
\mp {\hbar \over 2} \nabla_{j}[\nabla \times ({\bf V} \mp {\bf V}_{q})]
= 0\,,\eqno(4.11)$$
which is identically satisfied by Eq.(4.7).
Deriving this equation we explored the identity
$${\hbar^2\over 2m}{\nabla \sqrt{\rho} \over \sqrt{\rho}}
= {m {\bf V}^2_{q}\over 2} - {\hbar \over 2}[\partial_{1} ({\bf V}_{q})_{2}
- \partial_{2} ({\bf V}_{q})_{1}],$$
and the Chern-Simons Gauss law (3.15).
Thus, under condition (4.7)
it remains to satisfy only the vorticity condition (3.16)
for the quantum velocity
$$\nabla \times {\bf V}_{q} =
\pm {e^2 \over \kappa m c^2}\,\rho\,,\eqno(4.12)$$
acquiring by definition (4.4) Liouville's equation form
$$\Delta \ln \rho = \mp {2 e^2 \over \kappa \hbar c^2}\, \rho.\eqno(4.13)$$
We stress again that the Liouville equation in our model has
meaning of the vorticity condition for quantum flow.
Solutions of the model are well known [13,15]. We just mention
the polar symmetric case (for the sign minus)
$$\rho = 4 {\kappa \hbar c^2 N^2 \over e^2 r^2 } [({r \over r_{0}})^N +
({r_{0}\over r})^N]^{-2}\,,
\eqno(4.14)$$
which is regular for $N \ge 1$ and appears from the general solution
$$\rho = \alpha {|\zeta'(z)|^2 \over (1 + |\zeta (z)|^2)^2}\,,\eqno(4.15)$$
when
$$\zeta (z) = {c_{N} \over (z - z_{0})^N},
\,\,\,z = x + iy.\eqno(4.16)$$
\par
Now there are two physical conditions in the original
($\psi$, ${\bf A}$) formulation,  restricting our solution. From regularity of
the gauge potential ${\bf A}$ we fix the phase of $\chi = \sqrt{\rho}
\exp(iS/\hbar)$ (see Eqs.(3.6)) as
$S/\hbar = (N - 1)\theta$, $\theta = \tan^{-1} (x_{2}/x_{1})$, and restrict
$N$ to be an integer for single-valued $\chi$. But single-valuedness of the
original function $\psi = \sqrt{\rho}\exp (i S)$ in (3.5) requires integer valuedness
for the product
$$(N - 1)\hbar = integer,\eqno(4.17)$$
valuedness of which for any integer $N$ leads to an integer valuedness of
deformation parameter
$$\hbar = n\,,\eqno(4.18)$$
and as a consequence of (4.10), we find the quantization condition
$${\kappa g \over e^2} = \pm{n \over 2 mc^2}, (n = 1,2,3,...)\,.\eqno(4.19)$$
The last relation means that the Chern-Simons coupling constant and the
quantum potential strength must be quantized
$$ \kappa = n {e^2\over  2 g m c^2},\,\,\,\,\,\,
1 - \hbar^2 = 1 - n^2 = (1 - n)(1 + n).\eqno(4.20)$$
\par
At the end of this section we present the Lagrangian formulation of our
fluid model (3.16). After excluding the vector potentials $A_{\mu}$
from (3.7) we get
$$L = {\kappa m^2 c^2\over 2 e^2 }\epsilon_{\mu \nu \lambda } V_{\mu}
\partial_{\nu}V_{\lambda} - \rho V_{0} - \rho {m {\bf V}^2 \over 2}
- \rho {m {\bf V_{q}}^2 \over 2} + g \rho^2,\eqno(4.21) $$
where $V_{0}$ plays the role of Lagrange multiplier.
The Hamiltonian (constrained by Chern-Simons Gauss law)
$$H = \int  \rho {m {\bf V}^2 \over 2} +
\rho {m {\bf V_{q}}^2 \over 2} - g \rho^2,\eqno(4.22)$$
has simple interpretation as the sum of kinetic energies of the
classical and quantum motions, plus self-interaction energy.
As easy to check
it vanishes for the self-dual flow (4.7) with fixed constants (4.10).

\bigskip\par
\noindent
{\bf 5. 1+1 dimensional reduction}
\bigskip
For the one directional flow, assume in the $x$ direction when
$\partial_{2} = 0$,
the system (3.16) reduces to
$$\partial_{0} V_{1} + V_{1}\partial_{1} V_{1} = {1 \over m}
\partial_{1}(2 g \rho + {\hbar^2 \over 2 m}{\partial^2_{1}\sqrt{\rho}
\over \sqrt{\rho}}), \eqno(5.1a)$$
$$\partial_{0} V_{2} + V_{1} \partial_{1} V_{2} = 0,\eqno(5.1b)$$
$$\partial_{1} V_{2} = {e^2 \over \kappa m c^2}\rho.\eqno(5.1c)$$
Substituting $\partial_{1} V_{2}$ from the last equation to the
second one we find that velocity field component $V_{2}$
is decoupled completely from Eq.(5.1a) and is determined by the
first order system of  equations
$$\partial_{1} V_{2} = {e^2 \over \kappa m c^2}\rho,\,\,\,
\partial_{0} V_{2} = -{e^2 \over \kappa m c^2}\rho V_{1}.\eqno(5.2)$$
Then compatibility condition for the last system is just the
continuity equation for one dimensional flow
$$\partial_{0}\rho + \partial_{1}(\rho V_{1}) = 0\,. \eqno(5.3)$$
Equations (5.1a) and (5.3) determine the Madelung fluid in
one space dimension. Rewriting them for the wave function
$$\chi = \sqrt{\rho} e^{{i \over \hbar} \int_{-\infty}^{x}V_{1}}\,,\eqno(5.4)$$
we obtain the NLS model
$$i\hbar \partial_{0}\chi + {\hbar^2 \over 2m}\partial^2_{1}\chi
+ 2 g |\chi|^2 \chi = 0\,.\eqno(5.5)$$
\par
We note that for the negative sign of the quantum deformation in the system
(3.5), corresponding to the replacement
$\hbar^2 \rightarrow -\hbar^2$,
the one dimensional reduction of the fluid is given by the system
(5.3), (5.1a), where in the last equation we must change the
sign of the quantum potential contribution. The result is that
for the wave function
$$\psi = \sqrt{\rho} e^{i  \int_{-\infty}^{x}V_{1}}\,,\eqno(5.6)$$
we get NLS modified by quantum potential

 $$i\partial_{0}\psi + {1 \over 2m}\partial^2_{1}\psi
+ 2 g |\psi|^2 \psi = (1 + \hbar^2){1 \over 2m}
{\partial^2_{1}|\psi| \over |\psi|} \psi\,.
\eqno(5.7)$$
At the same time in terms of two real functions
$$Q^{\pm} = \sqrt{\rho} e^{\pm {1 \over \hbar}\int^{x}_{-\infty} V_{1}}\,,
\eqno(5.8)$$
we have "dissipative" (reaction-diffusion) version of NLS
$$\pm \hbar \partial_{0}Q^{\pm} + {\hbar^2 \over 2m}\partial^2_{1}
Q^{\pm} + 2g Q^{+}Q^{-}Q^{\pm}  = 0\,.\eqno(5.9)$$
This analogy allow us derive bilinear representation for
(5.7). Solution for the wave function $\psi$ is represented
in terms of three real
functions $G^{\pm}, F$,
$$\psi = {(G^+)^{1-i\hbar \over 2} (G^-)^{1+i\hbar \over 2} \over F},\,\,\,
\bar\psi = {(G^+)^{1+i\hbar \over 2} (G^-)^{1-i\hbar \over 2} \over F}
\eqno(5.9)$$
satisfying bilinear system of equations
$$(\pm \hbar D_{t} -
{\hbar^2 \over 2m}D^2_{x})(G^{\pm} * F) = 0\,,\eqno(5.11a)$$
$${\hbar^2 \over 2m}D^2_{x}(F * F) = 2g G^+ G^-\,.\eqno(5.11b)$$
Then, hydrodynamical variables are given by formulae
$$V_{1} = {\hbar \over 2m} \partial_{1}\ln {G^- \over G^+},
\,\,\,
\rho = {\hbar^2 \over 2m g} \partial^2_{1}(\ln F)\,.\eqno(5.12)$$
Constructing one and two-soliton solutions we find that in contrast to
the NLS case, soliton dynamics of Eq.(5.7)
have a reach resonance phenomenology
[4,8]. Another reduction of 2+1 dimensional model  (3.5)
leads to the DNLS type
equation and its reaction-diffusion analog. Details of reduction procedure
and resonance soliton interactions would be published elsewhere.
\bigskip\par
\noindent
{\bf 6. Conclusions}
\bigskip\par
We reformulated the classical dynamics of non-relativistic particle
interacting with Abelian gauge field as a nonlinear wave equation
with quantum potential.
Then we considered deformations of this equation and found two cases
depending of the sign of deformation. For one of the signs the
standard Schr\"odinger model with deformation parameter playing the role
of the Planck constant was obtained. While for the second sign 
we got diffusion-anti-diffusion equation. Specifying the gauge field
as Chern-Simons one and including cubic nonlinear term to the Schr\"odinger
equation we found the dispersionless limit of the Jackiw-Pi  model
which could be useful descriptive of the anyon's  semiclassical limit.
The deformation of this model is equivalent to the standard JP model
which we represented as a rotational hydrodynamics of Madelung type
fluid. Special flow in
this fluid, when velocities of the classical and quantum motion coincide,
lead to the Liouville equation admitting vortex configurations.
The similar equation, as we found before [4,8], defines the event
horizon for black hole type solution in the one dimensional NLS with quantum
potential (5.7). Moreover,
in terms of the wave function it is exactly the Chern-Simons
self-(anti-self)duality condition [13]. In fact the self-duality equations
are first order equations why they can be interpreted in terms of
velocity fields and
we hope that our interpretation could be applied to other models as well.

\bigskip
\par\noindent{\bf Acknowledgments}
\bigskip

One of the authors would like to thank
Professor Pierre Sabatier for useful discussion and
Professor Fon-Che Liu and Institute of
Mathematics, Academia Sinica, Taipei for warm hospitality. This work was
supported in part by the Izmir Institute of Technology, Turkey and the
Institute of Mathematics, Academia Sinica, Taipei, Taiwan.
\bigskip
\par\noindent
{\bf References}
\bigskip
\par\noindent
[1] Guerra F., Pusterla M. {\it A nonlinear Schr\"odinger equation and
its relativistic generalization from basic principles}, Lett. Nuovo
Cimento, 34 (1982) 351-356; Visier J.P., {\it Particular solutions of a
non-linear Schr\"odinger equation carrying particle-like singularities
represent possible models of de Broglie's double solution theory},
Phys. Lett. A 135 (1989) 99-105.
\par\noindent
[2] Smolin L.,{\it Quantum fluctuations and inertia}, Phys. Lett. A 113
(1986) 408-12; Bertolami O.,{\it Nonlinear corrections to quantum
mechanics from quantum gravity}, Phys. Lett. A154 (1991) 225-9.
\par\noindent
[3] Schiller R., {\it Quasi-Classical Theory of the Nonspinning Electron},
Phys. Rev. 125 (1962) 1100-1108; 1109-1115; 1116-1123;
Rosen N., {\it The relation between classical and quantum mechanics},
Amer. J. Phys. 32 (1964) 597-600.
\par\noindent
[4] Pashaev O.K. and Lee J.H., {\it Black holes and solitons of
quantized dispersionless NLS and DNLS equations}, ANZIAM Journal of Appl. Math,
2001 (in press).
\par\noindent
[5] Sabatier P.C., {\it Multidimensional nonlinear Schr\"odinger
equations with exponentially confined solutions}, Inverse
Problems, 6 (1990) L47-53.
\par\noindent
[6] Auberson G. and Sabatier P.C., {\it On a class of homogeneous
nonlinear Schr\"odinger equations}, J. Math. Phys. 35 (1994)
4028-40.
\par\noindent
[7] Bohm D., {\it A suggested interpretation of the quantum theory
in terms of hidden variables I}, Phys. Rev. 85 (1952) 166-79.
\par\noindent
[8] Pashaev O.K. and Lee J.H., {\it Soliton resonances, black holes
and Madelung fluid}, J. Nonlinear Math. Phys. 8, Supplement, 1-5,
(2001) (in press); {\it Resonance NLS solitons as black holes in
Madelung fluid}, hep-th/9810139.
\par\noindent
[9] Martina L, Pashaev O.K. and Soliani G., {\it Integrable dissipative
structures in the gauge theory of gravity}, Class. Quantum Grav.
14 (1997) 3179-86; {\it Bright solitons as black holes}, Phys. Rev.
D 58 (1998) 084025.
\par\noindent
[10] Schr\"odinger E., {\it \"Uber die Umkehrung der Naturgesetze},
Sitzungsberichte der Preussischen Akad. der Wissenschaft
Physicalisch Mathematische Klasse, (1931) 144-153.
\par\noindent
[11] Jin S., Levermore C.D. and McLaughlin D.W., Comm. Pure Appl.
Math. 52 (1999) 613-654;
\par\noindent
[12] Ercolani N and Montgomery R., {\it On the fluid approximation to
a nonlinear Schr\"odinger equation}, Phys. Lett. A180 (1993) 402-408.
\par\noindent
[13] Jackiw R. and Pi S-Y, Phys. Rev. Lett., 64 (1990) 2969;
{\it Classical and quantal nonrelativistic Chern-Simons theory},
Phys. Rev. D 42 (1990) 3500-13.
\par\noindent
[14] Salesi G, Mod. Phys. Lett. A22 (1996) 1815;
Salesi G. and Recami E.,{\it Hydrodynamics of spinning particles},
hep-th/9802106.
\par\noindent
[15] Arkadiev V. A., Pogrebkov A.K. and Polivanov M.C., Inverse Problems,
5 (1989) L1.
\end